\documentclass[aps,prb,reprint,showpacs]{revtex4-1}
\usepackage{graphicx} 
\usepackage{tabularx}
\usepackage{dcolumn} 
\usepackage{color}
\usepackage{bm}
\usepackage{amsmath}
\usepackage{amssymb}
\newcommand{\bi}[1]{\ensuremath{\boldsymbol{#1}}} 
\newcommand{\nn}{\nonumber}
 
\begin{document} 
 
\title{ 
Majorana surface states of superfluid $^3$He A- and B-phases in a slab
}

\author{Y. Tsutsumi} 
\affiliation{Department of Physics, Okayama University, 
Okayama 700-8530, Japan} 
\author{M.  Ichioka} 
\affiliation{Department of Physics, Okayama University, 
Okayama 700-8530, Japan} 
\author{K.  Machida} 
\affiliation{Department of Physics, Okayama University, 
Okayama 700-8530, Japan} 
\date{\today}

\begin{abstract} 
Motivated by experiments on the superfluid $^3$He confined in a thin slab,
we design a concrete experimental setup for observing the Majorana surface states.
We solve the quasi-classical Eilenberger equation, which is quantitatively reliable, to evaluate several quantities,
such as local density of states (LDOS), mass current for the A-phase, and spin current for the B-phase.
In connection with realistic slab samples,
we consider the upper and lower surfaces and the side edges including the corners with several thicknesses.
Consequently the influence on the Majorana zero modes from the spatial variation of $l$-vector for the A-phase in thick slabs
and the energy splitting of the zero-energy quasi-particles for the B-phase confined in thin slabs are demonstrated.
The corner of slabs in the B-phase is accompanied by the unique zero-energy LDOS of corner modes.
On the basis of the quantitative calculation, 
we propose several feasible and verifiable experiments to check the existence of the Majorana surface states,
such as the measurement of specific heat, edge current, and anisotropic spin susceptibility.
\end{abstract} 
 
 
%67.30.H- Superfluid phase of 3He   
%67.30.he Textures and vortices   
%74.20.-z Theories and models of superconducting state   
%74.20.Rp Pairing symmetries (other than s-wave)   
%74.25.-q Properties of type I and type II superconductors   
%74.25.Qt Vortex lattices, flux pinning, flux creep   
% 67.30.ht	Restricted geometries
%74.45.+c 	Proximity effects; Andreev reflection; SN and SNS junctions
 
\maketitle

\section{Introduction}

Majorana quasi-particles (QPs) and Majorana fermions have  attracted much attention in the wide research field, 
ranging from high energy physics to low temperature physics of ultracold atomic gases,~\cite{mizushima:2008, tsutsumi:2010}
and for the application to topological quantum computations.~\cite{nayak:2008}
The Majorana QP and Majorana fermionic operator  are defined by $\gamma^{\dagger }=\gamma$ and  $\Psi^{\dagger }=\Psi$, respectively,
which imply that the particle and antiparticle are identical.
It has been proposed that the Majorana nature brings new physics, 
such as non-abelian statistics of vortices in chiral superfluids~\cite{ivanov:2001} 
and Ising-like spin dynamics for time-reversal invariant superfluids.~\cite{chung:2009, volovik:2009a, nagato:2009}
The Majorana nature itself is an intriguing subject to further studies.

Candidate systems that exhibit the Majorana nature are quite rare, e.g., spin-triplet superconductors or superfluids,
$p$-wave Feshbach resonated superfluids,~\cite{mizushima:2008, tsutsumi:2010}
fractional quantum Hall systems with the 5/2 filling,~\cite{moore:1991, read:2000}
interfaces between a topological insulator and an $s$-wave superconductor,~\cite{fu:2008}
and $s$-wave superfluids with particular spin-orbit interactions.~\cite{sato:2009}

Spin-triplet superconductors or superfluids satisfy the following Bogoliubov-de Gennes equation with the wave function $\varphi_n$ and the energy $E_n$:
\begin{align}
\int d\bi{r}_2\widehat{\mathcal{H}}(\bi{r}_1,\bi{r}_2)\varphi_n(\bi{r}_2)=E_n\varphi_n(\bi{r}_1).
\end{align}
The eigenstates yield one-to-one mapping between the positive energy states $\varphi_E$ and the negative energy states $\varphi_{-E}=\widehat{\tau }_x\varphi_E^*$
owing to the symmetry $\widehat{\mathcal{H}}=-\widehat{\tau }_x\widehat{\mathcal{H}}^*\widehat{\tau }_x$,
where $\widehat{\tau }_x$ is the Pauli matrix in the particle-hole space.
The symmetry of the wave function leads to the relation of the Bogoliubov QP operator $\gamma_E=\gamma_{-E}^{\dagger }$.
Therefore, the Bogoliubov QP with zero-energy is the Majorana QP, $\gamma^{\dagger}_{0}=\gamma_{0}$.
Zero-energy bound states of the QPs appear whenever the underlying potential for the QPs changes its sign.
The situation appears at edges, surfaces, or vortices with odd integer winding numbers,
where the QPs have zero-energy Andreev bound states.
For the edge or surface, the Majorana fermion surface state
$\Psi^{\dagger }=\Psi$ exists as well as $\gamma^{\dagger}_{0}=\gamma_{0}$.~\cite{read:2000,stone:2004}

Among known possible spin-triplet superconductors~\cite{machida:1991,ohmi:1993,machida:1999,machida:2001} UPt$_3$, UGe$_2$, and URhGe,
Sr$_2$RuO$_4$ is a prime candidate of chiral spin-triplet superconductors.
The precise pairing symmetry, however, has not yet been identified and has been under strong discussion.
A $p_x + ip_y$ scenario is reexamined from various aspects.~\cite{machida:2008}
Obviously, we want to experiment with more candidate materials for observing the Majorana nature.

The superfluid $^3$He consists of spin-triplet Cooper pairs.~\cite{leggett:1975}
For the $^3$He there is a huge amount of experimental data
by intensive researches for long years.~\cite{vollhardt:book}
There are two stable phases, ABM- (A-) and BW- (B-) phases, for the superfluid $^3$He in a bulk without a magnetic field.
The A-phase is stabilized in narrow regions at high temperatures and high pressures 
while the B-phase is stabilized in other wide regions within the superfluid phase.
The order parameter (OP) of the A-phase has point nodes toward $l$-vector,
which signifies the direction of the orbital angular momentum, or orbital chirality, whereas the B-phase has a full gap.
The Majorana natures for the A- and B-phases differ due to the topology of the gap on the Fermi surface.~\cite{volovik:book}
Time reversal symmetry is also different for the A- and B-phases.
Since up-up and down-down spin Cooper pairs have the same chirality for the A-phase,
the time-reversal symmetry is broken and net mass current flows along the boundary or the edge.
In contrast, since up-up and down-down spin Cooper pairs have the opposite chirality, the B-phase is a time-reversal invariant.
Instead of the mass current canceled by up-up and down-down spin Cooper pairs, the spin current flows.
The difference of time reversal symmetry between the A- and B-phases 
is analogous to the difference between the quantum Hall state and the quantum spin Hall state.~\cite{qi:2009}

Our aim is to propose a concrete experimental design to detect the Majorana nature based on quantitative calculations
and understand systematically how to observe the Majorana nature in the superfluid $^3$He A- and B-phases.
The quantitative calculations are performed by quasi-classical theory which yields information of QPs.
The quasi-classical theory is valid when $\Delta / E_F \ll 1$, where $\Delta$ is superfluid gap and $E_F$ is the Fermi energy.
Since $\Delta / E_F \sim 10^{-3}$, the theory is appropriate for the superfluid $^3$He.~\cite{serene:1983}
The quasi-classical framework is well established within the weak-coupling limit.
The superfluid $^3$He at the lower pressure can be described within the weak-coupling scheme~\cite{greywall:1986}
without delicate strong-coupling corrections.

In the slab geometry, both the A- and B-phases are stabilized 
by changing thickness and temperature even at zero pressure.~\cite{li:1988, vorontsov:2003}
Therefore, we can deal with the A- and B-phases by the quasi-classical theory within the weak-coupling limit.
Here, we introduce several systems of the slab geometry.
By Bennett {\it et al.},~\cite{bennett:2010} the superfluid $^3$He is confined 
in a thin slab box with a thickness $D=0.6$ $\mu$m and an area of the base 10 mm $\times$ 7 mm.
The thickness is of the order of $10\xi_0$ at $P=0$, where $\xi_0$ is the coherence length at $T=0$.
At ISSP, University of Tokyo,~\cite{yamashita:2008} sample disks with a thickness $D=12.5$ $\mu$m, 
which is of the order of the dipole coherence length, and a diameter 3 mm are used.
The ISSP group can rotate the sample disks with the highest speed in the world and investigates half-quantum vortices.
At RIKEN,~\cite{saitoh:2007} the inter-digitated capacitors are used for making a film of the superfluid $^3$He.
The group can control thicknesses of the film from 0.3 to 4 $\mu$m.

\begin{figure}
\begin{center}
\includegraphics[width=8.5cm]{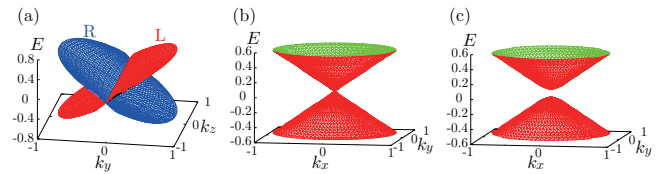}
\end{center}
\caption{(Color online) 
Stereographic views of typical dispersion relation.
(a) Dirac valley for the A-phase in a thin slab $D=8\xi_0$ at the left side edge ``L" and the right side edge ``R".
(b) Majorana cone for the B-phase in a thick slab $D=30\xi_0$ at the upper or lower surface.
(c) Split Majorana cone for the B-phase in a thin slab $D=14\xi_0$ at the upper or lower surface.
The units of energy and wave length are $\pi k_BT_c$ and $k_F$, respectively.
}
\label{3D}
\end{figure}

Under those experimental situations,
we consider the surface states from not only the upper and lower surfaces in a slab but also the side edges including the corners
and evaluate the dependence of the Majorana state on the thickness of a slab.
By the quasi-classical calculation, we show the``Dirac valley" (Fig.~\ref{3D}(a)) for the A-phase at the side edges.
In a thick slab, we consider the $k_z$-component of the OP, 
which was neglected in the previous work~\cite{tsutsumi:2010b} since $l$-vector was assumed to be perpendicular to the thin slab.
Several groups have demonstrated that the Majorana fermion surface state has the dispersion of $E=(\Delta/k_F)\left|\bi{k}_{\parallel }\right|$ 
with the surface perpendicular to the $z$-axis in the B-phase,
where $k_F$ is the Fermi wave number and $\bi{k}_{\parallel }=(k_x,k_y)$.~\cite{chung:2009,volovik:2009a,nagato:2009}
This implies that the surface state consists of a single Majorana cone.
We show the Majorana cone (Fig.~\ref{3D}(b)) by the quasi-classical calculation for a thick slab
and investigate the variation of the dispersion by the thickness of a slab (Fig.~\ref{3D}(c)).
The zero-energy state at the corner is also discussed.

The arrangement of this paper is as follows:
In Sec.~\ref{sec:theory}, we formulate the quasi-classical theory based on the quasi-classical Green's function 
which gives quantitative information of QPs.
We explain numerical methods in Sec.~\ref{sec:sys},
which is supplemented by the symmetry considerations of the quasi-classical Green's function given in Appendix.
In Secs.~\ref{sec:A} and \ref{sec:B} for the A- and B-phases, we show results of the spatial structures of the OP, the current distribution,
and the local density of states (LDOS) for QPs which relates to the Majorana modes.
In Sec.~\ref{sec:discuss}, we discuss difference of the Majorana zero modes in the A- and B-phase 
and propose several experimental designs to observe the Majorana nature.
In addition, we discuss the Majorana zero mode in the stripe phase.
The final section is devoted to a summary.

\section{\label{sec:theory}Quasi-Classical Theory and Order Parameters}

We start with the quasi-classical spinful Eilenberger equation,~\cite{eilenberger:1968}
which has been used for studies of the superfluid $^3$He.~\cite{schopohl:1980, fogelstrom:1995, vorontsov:2003}
The low energy excitation modes at the surface are discretized in the order of $\Delta/(k_FL)$,~\cite{stone:2004} where $L$ is a length of the system.
If $L$ is a macroscopic length, which is much larger than $k_F^{-1}$, the low energy excitation modes at the surface can be regarded as continuous spectra.
Because there is the exact zero-energy excitation at the surface of the superfluid $^3$He A-phase~\cite{stone:2004,tsutsumi:2010b} and B-phase,~\cite{chung:2009, volovik:2009a, nagato:2009}
the quasi-classical theory can be used to discuss the Majorana QP.

The quasi-classical Green's function $\widehat{g}(\bi{k},\bi{r},\omega_n)$ is calculated using the Eilenberger equation
\begin{multline}
-i\hbar\bi{v}(\bi{k})\cdot\bi{\nabla }\widehat{g}(\bi{k},\bi{r},\omega_n) \\
= \left[
\begin{pmatrix}
i\omega_n\hat{1} & -\hat{\Delta }(\bi{k},\bi{r}) \\
\hat{\Delta }(\bi{k},\bi{r})^{\dagger } & -i\omega_n\hat{1}
\end{pmatrix}
,\widehat{g}(\bi{k},\bi{r},\omega_n) \right],
\label{Eilenberger eq}
\end{multline}
where the ``ordinary hat" indicates the 2 $\times$ 2 matrix in spin space 
and the ``wide hat" indicates the 4 $\times$ 4 matrix in particle-hole and spin spaces.
The quasi-classical Green's function is described in particle-hole space by
\begin{align}
\widehat{g}(\bi{k},\bi{r},\omega_n) = -i\pi
\begin{pmatrix}
\hat{g}(\bi{k},\bi{r},\omega_n) & i\hat{f}(\bi{k},\bi{r},\omega_n) \\
-i\underline{\hat{f}}(\bi{k},\bi{r},\omega_n) & -\underline{\hat{g}}(\bi{k},\bi{r},\omega_n)
\end{pmatrix},
\end{align}
with the direction of the relative momentum of a Cooper pair $\bi{k}$, the center-of-mass coordinate of a Cooper pair $\bi{r}$,
and the Matsubara frequency $\omega_n=(2n+1)\pi k_B T$.
The quasi-classical Green's function satisfies a normalization condition $\widehat{g}^2=-\pi^2\widehat{1}$.
The Fermi velocity is given as $\bi{v}(\bi{k})=v_F\bi{k}$ on the three dimensional Fermi sphere.

We solve Eq.~\eqref{Eilenberger eq} by the Riccati method.~\cite{schopohl:1995,nagato:1993,vorontsov:2003}
We introduce Riccati amplitude $\hat{a}=(\hat{1}+\hat{g})^{-1}\hat{f}$ 
and $\hat{b}=(\hat{1}+\underline{\hat{g}})^{-1}\underline{\hat{f}}$
related to particle- and hole-like projections of the off-diagonal propagators, respectively.
Equation \eqref{Eilenberger eq} can be rewritten as Riccati equations
\begin{multline}
\hbar\bi{v}(\bi{k})\cdot\bi{\nabla }\hat{a}(\bi{k},\bi{r},\omega_n) \\
= \hat{\Delta }(\bi{k},\bi{r})-\hat{a}(\bi{k},\bi{r},\omega_n)\hat{\Delta }(\bi{k},\bi{r})^{\dagger } \hat{a}(\bi{k},\bi{r},\omega_n)\\
-2\omega_n\hat{a}(\bi{k},\bi{r},\omega_n), \nn
\end{multline}
\begin{multline}
-\hbar\bi{v}(\bi{k})\cdot\bi{\nabla }\hat{b}(\bi{k},\bi{r},\omega_n) \\
= \hat{\Delta }(\bi{k},\bi{r})^{\dagger }-\hat{b}(\bi{k},\bi{r},\omega_n)\hat{\Delta }(\bi{k},\bi{r})\hat{b}(\bi{k},\bi{r},\omega_n)\\
-2\omega_n\hat{b}(\bi{k},\bi{r},\omega_n).
\label{Riccati eqs}
\end{multline}
The equations are solved by integration toward $\bi{k}$ for $\hat{a}(\bi{k},\bi{r},\omega_n)$
and toward $-\bi{k}$ for $\hat{b}(\bi{k},\bi{r},\omega_n)$.
From the Riccati amplitude, the quasi-classical Green's function is given as
\begin{align}
\widehat{g} = -i\pi
\begin{pmatrix}
(\hat{1}+\hat{a}\hat{b})^{-1} & 0 \\
0 & (\hat{1}+\hat{b}\hat{a})^{-1}
\end{pmatrix}
\begin{pmatrix}
\hat{1}-\hat{a}\hat{b} & 2i\hat{a} \\
-2i\hat{b} & -(\hat{1}-\hat{b}\hat{a})
\end{pmatrix}.
\end{align}

The self-consistent condition for the pair potential $\hat{\Delta }(\bi{k},\bi{r})$ is given as
\begin{align}
\hat{\Delta }(\bi{k},\bi{r}) = 
N_0\pi k_BT\sum_{-\omega_c \le \omega_n \le \omega_c}\left\langle V(\bi{k}, \bi{k}') \hat{f}(\bi{k}',\bi{r},\omega_n)\right\rangle_{\bi{k}'},
\end{align}
where $N_0$ is the density of states in the normal state, 
$\omega_c$ is a cutoff energy setting $\omega_c=40\pi k_B T_c$ with the transition temperature $T_c$ in a bulk,
and $\langle\cdots\rangle_{\bi{k}}$ indicates the Fermi surface average.
The pairing interaction $V(\bi{k}, \bi{k}')=3g_1\bi{k}\cdot\bi{k}'$ 
for Cooper pairs with an orbital angular momentum $L=1$, where $g_1$ is a coupling constant.
In our calculation, we use a relation
\begin{align}
\frac{1}{g_1N_0}=\ln \frac{T}{T_c}+2\pi k_BT\sum_{0 \le \omega_n \le \omega_c}\frac{1}{\omega_n}.
\end{align}

Spin-triplet OP is defined by a vectorial notation
\begin{align}
\hat{\Delta }(\bi{k},\bi{r})=(i\hat{\bi{\sigma }}\hat{\sigma }_y)\cdot \bi{\Delta }(\bi{k},\bi{r}),
\end{align}
with the Pauli matrix $\hat{\bi{\sigma }}$ in spin space.
The complex vector $\bi{\Delta }$ is perpendicular to the spin $\bi{S}$ of a Cooper pair, namely, $\bi{\Delta }\cdot\bi{S}=0$.
The $\bi{\Delta }$ can be expanded in orbital momentum directions,
\begin{align}
\Delta_{\mu }(\bi{k},\bi{r})=A_{\mu i}(\bi{r})k_i,
\end{align}
where $A_{\mu i}(\bi{r})$ is a complex 3 $\times$ 3 matrix with a spin index $\mu$ and an orbital index $i$.
The repeated index implies summation over $x$, $y$, and $z$.

This paper is discussed for the chiral state in the A-phase, polar state, B-phase, and planar state.
Symbolic descriptions of the OP in their state are the following:
For the chiral state,
\begin{align}
\Delta=d_z(k_x+ik_y),
\end{align}
for the polar state,
\begin{align}
\Delta=d_zk_x,
\end{align}
for the B-phase,
\begin{align}
\Delta=d_xk_x+d_yk_y+d_zk_z,
\end{align}
and for the planar state,
\begin{align}
\Delta=d_xk_x+d_yk_y,
\end{align}
where $d$-vector is perpendicular to the spin of a Cooper pair.
$d$-vector and the momentum direction are permitted to rotate spherically in the spin and momentum space, respectively.

By using the self-consistent quasi-classical Green's function, the mass and spin currents are calculated by
\begin{align}
\bi{j}(\bi{r}) &=
mN_0\pi k_BT\sum_{-\omega_c \le \omega_n \le \omega_c} 
\langle \bi{v}(\bi{k}) \ {\rm Im} \left[ g_0(\bi{k},\bi{r}, \omega_n) \right] \rangle_{\bi{k}}, 
\label{mass current}\\
\bi{j}_s^{\mu }(\bi{r}) &=
\frac{\hbar }{2}N_0\pi k_BT\sum_{-\omega_c \le \omega_n \le \omega_c} 
\langle \bi{v}(\bi{k}) \ {\rm Im} \left[ g_{\mu }(\bi{k},\bi{r}, \omega_n) \right] \rangle_{\bi{k}},
\label{spin current}
\end{align}
respectively, where $m$ is the mass of the $^3$He atom 
and $g_{\mu }$ is a component of the quasi-classical Green's function $\hat{g}$ in spin space, namely,
\begin{align}
\hat{g} =
\begin{pmatrix}
g_0+g_z & g_x-ig_y \\
g_x+ig_y & g_0-g_z
\end{pmatrix}. \nn
\end{align}

LDOS for energy $E$ is given by
\begin{align}
N(\bi{r},E) &= \left\langle N(\bi{r},E,\bi{k})\right\rangle_{\bi{k}} \nn\\
&= N_0 \left\langle {\rm Re} \left[g_0(\bi{k},\bi{r}, \omega_n)|_{i\omega_n \rightarrow E+i\eta}\right] \right\rangle_{\bi{k}},
\end{align}
where $\eta$ is a positive infinitesimal constant and $N(\bi{r},E,\bi{k})$ is angle-resolved LDOS.
Typically, we choose $\eta=0.003\pi k_BT_c$.
For obtaining $g_0(\bi{k},\bi{r}, \omega_n)|_{i\omega_n \rightarrow E+i\eta}$, 
we solve Eqs.~\eqref{Riccati eqs} with $\eta -iE$ instead of $\omega_n$ under the pair potential obtained self-consistently.

\section{\label{sec:sys}System Geometry and Numerical Methods}

\begin{figure}
\begin{center}
\includegraphics[width=8.5cm]{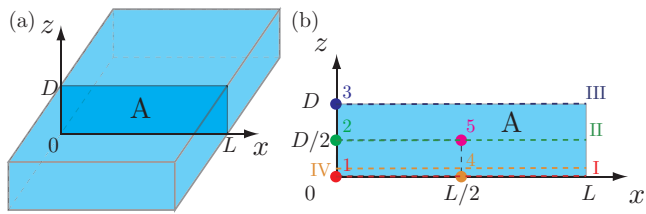}
\end{center}
\caption{(Color online) 
(a) A schematic configuration of a slab 
with a thickness of $D$ along the $z$-direction and a macroscopic length of $L$ along the $x$-direction.
We consider a cross-section A $(0\le x \le L,0\le z \le D)$.
(b) Indexes of positions in the cross-section A.
We label $(x,z)=(0,0)$, $(0,D/2)$, $(0,D)$, $(L/2,0)$, and $(L/2,D/2)$ as ``1", ``2", ``3", ``4", and ``5".
Paths $z=0$, $z=D/2$, $z=D$, and $z=0.5\xi_0$ along the $x$-axis are labeled as ``I", ``II", ``III", and ``IV".}
\label{system}
\end{figure}

We consider a cross-section A of a  slab 
with a thickness of $D$ along the $z$-direction and a macroscopic length of $L$ along the $x$-direction, shown in Fig.~\ref{system}(a).
It is assumed that the quasi-classical Green's function and OP are homogeneous along the $y$-direction.
We also assume that the surfaces of the slab are specular, where specularity is controlled by coating the surface with $^4$He atoms.~\cite{wada:2008}
Under the boundary condition, the quasi-classical Green's function or the Riccati amplitude 
changes only the direction of the relative momentum by mirror reflection at a surface $\bi{R}_{\rm surf}$, namely,
$\widehat{g}(\bi{k},\bi{R}_{\rm surf},\omega_n)=\widehat{g}(\underline{\bi{k}},\bi{R}_{\rm surf},\omega_n)$ or
$\hat{a}(\bi{k},\bi{R}_{\rm surf},\omega_n)=\hat{a}(\underline{\bi{k}},\bi{R}_{\rm surf},\omega_n)$ and
$\hat{b}(\bi{k},\bi{R}_{\rm surf},\omega_n)=\hat{b}(\underline{\bi{k}},\bi{R}_{\rm surf},\omega_n)$,
where $\underline{\bi{k}}=\bi{k}-2\bi{n}(\bi{n}\cdot\bi{k})$ with a unit vector $\bi{n}$ which is perpendicular to the surface.

We solve Eqs.~\eqref{Riccati eqs} by the numerical integration 
toward the $\bi{k}$-direction for $\hat{a}$ and toward the $-\bi{k}$-direction for $\hat{b}$ from the surface to the position $\bi{r}$.
Since $\hat{a}(\bi{k},\bi{R}_{\rm surf},\omega_n)=\hat{a}(\underline{\bi{k}},\bi{R}_{\rm surf},\omega_n)$,
the initial value of $\hat{a}$ at a surface $\bi{R}_{\rm surf}$ is determined by the numerical integration toward the $\underline{\bi{k}}$-direction 
from the other surface to $\bi{R}_{\rm surf}$.
Eventually, we have to solve the numerical integration along the sufficiently long path by changing the direction from $\bi{k}$ to $\underline{\bi{k}}$
at the surface so that the integration path gives the same value of $\hat{a}$ under arbitrary initial values, and also $\hat{b}$.
Since the integration path changes the direction at the surface, simple calculation is difficult.
For overcoming the difficulty, we exploit that the change in the direction of the integration path from $\bi{k}$ to $\underline{\bi{k}}$
can be regarded as the change in the relative momentum of the pair potential from $\bi{k}$ to $\underline{\bi{k}}$.
For example, we can substitute $-A_{\mu x}$ for $A_{\mu x}$ in the pair potential 
instead of the reflection of the integration paths at a surface $x=0$.
Similarly for the other surfaces, we can integrate Eqs.~\eqref{Riccati eqs} along the straight long paths in the region 
connected to the antiperiodic pair potential infinitely.

We calculate the quasi-classical Green's function in a range $-L/2 \le x \le L/2$ and $-D/2 \le z \le D/2$
instead of a cross-section A $(0\le x \le L,0\le z \le D)$ to reduce computational time.
We show the reduction method in Appendix 
in addition to other reduction methods with the symmetry of the quasi-classical Green's function.

\section{\label{sec:A}A-phase}

We discuss here the situation in which the thickness of a slab $D$ is changed,
where the A- and B-phases are stable in thin and thick slabs, respectively.~\cite{vorontsov:2003}
We present the calculated results of OP, LDOS, and mass current for the A-phase and spin current for the B-phase at $T=0.2T_c$.
In each phase, the results depend on the thickness $D$.
The length of the cross-section A is taken as $L=40\xi_0$ throughout the paper where the OP is recovered to the bulk value at the center of the system.
Note that the following results are unchanged for the length $L$ longer than $40\xi_0$.
In addition, we consider the low pressure limit $P\rightarrow 0$ within the weak-coupling limit.
The spatial positions of the results are indicated by the indexes in Fig.~\ref{system}(b).
In this section and next section, we use the units $\xi_0=\hbar v_F/2\pi k_BT_c$, $\pi k_BT_c$, and $N_0$ 
for length, energy, and LDOS, respectively.

The bulk of the A-phase is in the chiral state.
The OP in the chiral state is described by~\cite{vollhardt:book}
\begin{align*}
A_{\mu i}=d_{\mu }(\bi{m}+i\bi{n})_i,
\end{align*}
where $m$-vector and $n$-vector are perpendicular to each other.
$l$-vector, which signifies the orbital chirality, is defined as $\bi{l}\equiv\bi{m}\times\bi{n}$.
At the edge of a slab, since the normal orbital component to the edge vanishes,
the polar state will be realized.
The OP in the polar state is described by~\cite{vollhardt:book}
\begin{align*}
A_{\mu i}=d_{\mu }m_i,
\end{align*}
where we regard $n$-vector as perpendicular to the edge.

In a slab cell used by Bennett {\it et al.}~\cite{bennett:2010} where the thickness is 0.6 $\mu$m $\approx 8\xi_0$, 
it is expected that the A-phase is stable at $T=0.2T_c$.~\cite{vorontsov:2007}
Since the thickness $D$ is much shorter than the dipole coherence length $\sim 1000\xi_0$,
$d$-vector, which characterizes the spin state of the OP in the A-phase, points to the $z$-direction
in the absence of a magnetic field as long as $l$-vector is parallel to the $z$-axis.
$l$-vector is parallel to the $z$-axis everywhere except near the side edges at $x=0$ and $x=L$ in the slab shown in Fig.~\ref{system}.
The length scale of the spatial variation of $l$-vector is short, whose order is the coherence length.
On the other hand, $d$-vector can not vary spatially since the order of the length scale is the dipole coherence length.
$d$-vector is spatially uniform even under a magnetic field.~\cite{fetter:1976}
In this paper, we fix the direction of $d$-vector to the $z$-axis.

\subsection{$D=8\xi_0$: Thin slab for A-phase}

\begin{figure*}
\begin{center}
\includegraphics[width=17.5cm]{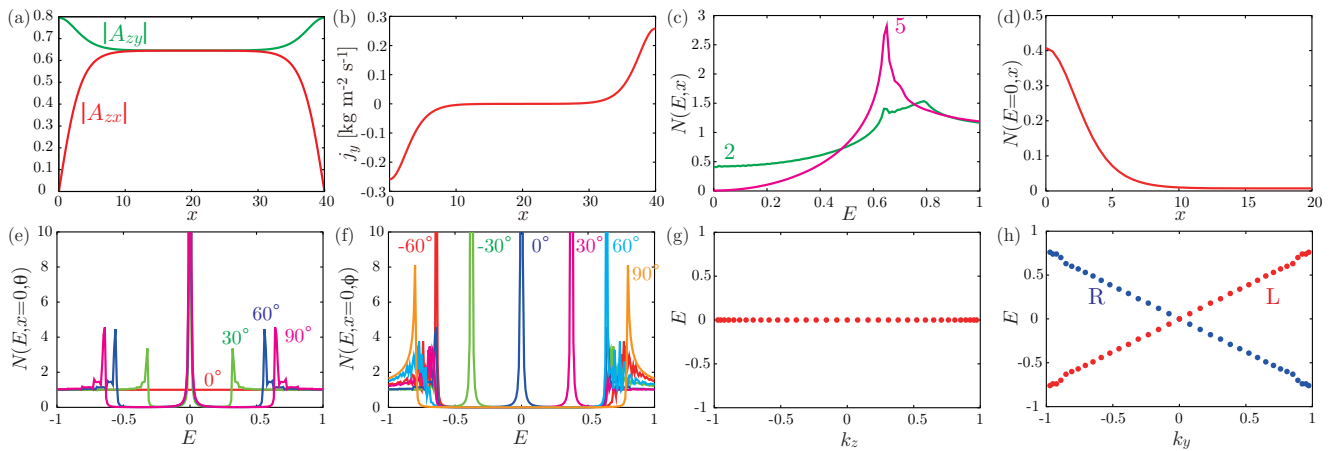}
\end{center}
\caption{(Color online) 
Calculated results for the A-phase with $D=8$.
Profiles of OP (a), and mass current $j_y$ (b) along the $x$-axis.
(c) LDOS $N(E,x)$ at ``2", and ``5" defined in Fig.~\ref{system}(b).
(d) Zero-energy LDOS $N(E=0,x)$ from the edge at $x=0$ to the bulk at $x=L/2$.
Angle-resolved LDOS $N(E,x=0,\theta)$ with $\phi=0^{\circ }$ (e),
and $N(E,x=0,\phi)$ with $\theta=90^{\circ }$ (f).
The peak energy of angle-resolved LDOS at the side edges as a function of $k_z$ for $k_y=0$ (g), and as a function of $k_y$ for $k_z=0$ (h),
where ``L" (``R") is for the left (right) edge.
In this and the following figures, the units of length, energy and $A_{\mu i}$, LDOS, and wave length are $\xi_0$, $\pi k_BT_c$, $N_0$, and $k_F$, respectively.}
\label{As}
\end{figure*}

For $D=8$, the OP is described by $\Delta_z(\bi{k},x)=A_{zx}(x)k_x+A_{zy}(x)k_y$,
where the relative phase between $A_{zx}$ and $A_{zy}$ is $\pi/2$.
The OP is uniform along the $z$-direction and varies along only the $x$-direction.
Since the slab is thin, the $k_z$-component of the OP is suppressed.
Note that the uniformness along the $z$-direction and the suppression of the $k_z$-component 
are not suppositions like in the previous work,~\cite{tsutsumi:2010b} but results by the self-consistent calculation.
The profile of the OP along the $x$-axis is shown in Fig.~\ref{As}(a).
Because of the specular boundary condition,
the $k_x$-component perpendicular to the edge becomes zero at $x=0$ and $x=L$.
In contrast, the parallel $k_y$-component is enhanced by compensating for the loss of the $k_x$-component at the edge,
where the polar state is realized.
The $k_x$-component increases and the $k_y$-component decreases toward the bulk region around $x=L/2$;
thus, the chiral state with $k_x+ik_y$ is attained.
We can construct $l$-vector as $l_i\equiv -i\epsilon_{ijk}A_{zj}^*A_{zk}/|\Delta|^2$,
where $\epsilon_{ijk}$ is the totally antisymmetric tensor and $|\Delta|^2=A_{zi}^*A_{zi}$ is the squared amplitude of the OP.
$l$-vector points to the $z$-direction in the bulk region and vanishes at the edge.
Since the chiral state is realized except at the edge, 
the Majorana fermion edge state $\Psi^{\dagger }=\Psi$ exists at side edges.~\cite{read:2000,stone:2004}

The mass current $j_y(x)$ is shown in Fig.~\ref{As}(b).
The mass current flows circularly along the side edge of the slab.
Experimental values~\cite{greywall:1986} are used as coefficients of Eq.~\eqref{mass current}
so that a quantitative value of the mass current is obtained.
By applying a magnetic field perpendicular to a slab,
we can produce a spin imbalance due to the Zeeman shift between the up-up spin pairs and the down-down spin pairs.
This results in a net spin current in addition to the mass current.

Figure~\ref{As}(c) shows LDOS at the edge $x=0$ (the position ``2" in Fig.~\ref{system}) and the bulk $x=L/2$ (``5").
It is clearly seen from the line ``2" for $x=0$
that the LDOS with a substantial weight appears at a zero-energy, corresponding to the Majorana edge mode,
because the distance $L$ between the edges is macroscopic.~\cite{read:2000,stone:2004}
This implies that the LDOS is expressed as $N(E, x=0)=N(E=0,x=0)+\alpha E^2$ in the vicinity of $E=0$.
The first term comes from the Majorana edge mode,
and the second term comes from the point node of the chiral state in the bulk A-phase.
The peak at $E\approx 0.65$ comes from the gap of the chiral state in the bulk
and the peak at $E\approx 0.8$ comes from the gap of the polar state at the edge.
The LDOS at $x=L/2$ (the line ``5") shows a typical behavior of the point node spectrum, namely, $N(E, x=L/2)\propto E^2$.
In Fig.~\ref{As}(d) we show the extent of the Majorana edge mode at $E=0$ toward the bulk from the edge at $x=0$, 
which spreads over the order of $5\xi_0$.
The spectrum of the edge gradually changes into the bulk spectrum.

The LDOS $N(\bi{r},E)$ is obtained by averaging the angle-resolved LDOS $N(\bi{r},E,\bi{k})$ on the Fermi surface, 
which has the peak at the energy of the surface Andreev bound state.
The dispersion relation of the surface Andreev bound state can be evaluated by following the peak in each momentum direction.
The angle-resolved LDOS $N(E,x=0,\theta)$ with $\phi=0^{\circ }$ and $N(E,x=0,\phi)$ with $\theta=90^{\circ }$
are shown in Figs.~\ref{As}(e) and \ref{As}(f), respectively, 
where $\theta$ is the polar angle from the $k_z$-axis and $\phi$ is the azimuthal angle from the $k_x$-axis on the three-dimensional Fermi sphere.
The angles and momentum directions have the relations: $\tan\theta=\sqrt{k_x^2+k_y^2}/k_z$ and $\tan\phi=k_y/k_x$.
The angle-resolved LDOS is symmetric between $0^{\circ }\le\theta\le 90^{\circ }$ and $180^{\circ }\ge\theta\ge 90^{\circ }$,
and between $-90^{\circ }\le\phi\le 90^{\circ }$ and $270^{\circ }\ge\phi\ge 90^{\circ }$.
The zero-energy peaks of the angle-resolved LDOS appear if $\phi=0^{\circ }$ except for $\theta=0^{\circ }$.
The zero-energy peaks come from the zero-energy Andreev bound state of QPs at the edge.
The QPs without the $k_y$-component have the zero-energy Andreev bound state
because they feel the $\pi$-phase shift of the pair potential 
for the relative momentum from $(k_x,k_y=0,k_z)$ to $(-k_x,k_y=0,k_z)$.
This is clear from the Eilenberger equation \eqref{Eilenberger eq}
since the quasi-classical Green's function $\widehat{g}(\bi{k})$ depends on only the $\bi{k}$-component of the pair potential $k_x+ik_y$.
The angle-resolved LDOS is asymmetric with respect to $E=0$ (Fig.~\ref{As}(f))
because the superfluid state in the A-phase has the chirality.~\cite{volovik:book}
The asymmetric angle-resolved LDOS is related to the mass current.

The peak energy of angle-resolved LDOS at the side edges as a function of $k_z$ for $k_y=0$ and as a function of $k_y$ for $k_z=0$ 
is shown in Figs.~\ref{As}(g) and \ref{As}(h), respectively.
It is found that the energy is dispersionless along the $k_z$-axis
and two linear dispersions appear near $E=0$ along the $k_y$-axis, 
where each dispersion comes from the left ($x=0$) and right ($x=L$) edges.
Since the directions of the mass current are different both edges,
the linear dispersions have the opposite slope.
The linear Dirac dispersions continue along the $k_z$-direction, which form a ``Dirac valley''.
The results are consistent 
with the strict quantum level structures of the Majorana fermion edge state by Bogoliubov-de Gennes equation.~\cite{tsutsumi:2010b}

\subsection{$D=14\xi_0$: Thick slab for A-phase}

\begin{figure}
\begin{center}
\includegraphics[width=8.5cm]{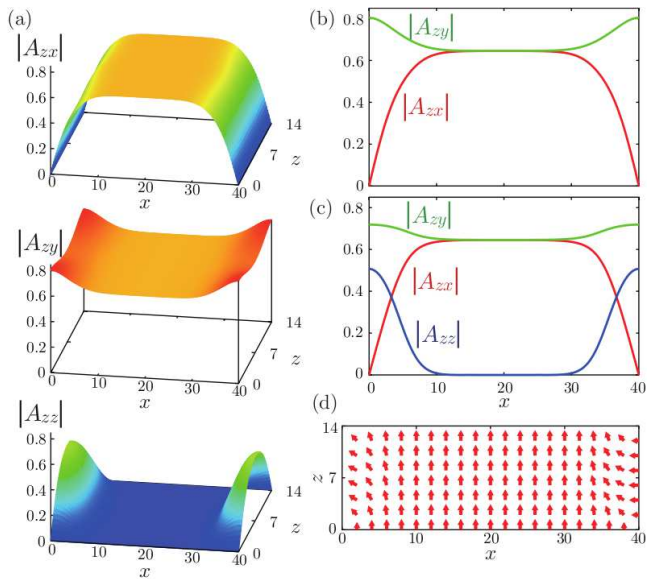}
\end{center}
\caption{(Color online) 
The OP for the A-phase with D=14.
(a) Amplitude of each component.
Profiles of each component along ``I" (b), and along``II" (c).
(d) Orientations of $l$-vector texture in the $x$-$z$ plane of a slab.}
\label{A-OP}
\end{figure}

In a thick slab, the $k_z$-component of the OP is induced.
For $D=14$, the $k_z$-component appears near the edges at $x=0$ and $x=L$ to avoid the polar state
except at $z=0$ and $z=D$, shown in Figs.~\ref{A-OP}(a)-\ref{A-OP}(c).
The OP is described by $\Delta_z(\bi{k},\bi{r})=A_{zx}(\bi{r})k_x+A_{zy}(\bi{r})k_y+A_{zz}(\bi{r})k_z$,
where the phases of $A_{zz}$ and $A_{zx}$ are the same 
and the relative phase between $A_{zz}$ and $A_{zy}$ is $\pi/2$.
Since the thickness of $D=14$ is small, the $A_{zz}$ does not recover the bulk value of $A_{zx}$
at the side edges $x=0$ and $x=L$ along $z=D/2$ (Fig.~\ref{A-OP}(c)).
As similarly at the edges $z=0$ and $z=D$, $l$-vector is perpendicular to the edges at $x=0$ and $x=L$ by the induced $k_z$-component,
shown in Fig.~\ref{A-OP}(d).

\begin{figure*}
\begin{center}
\includegraphics[width=17.5cm]{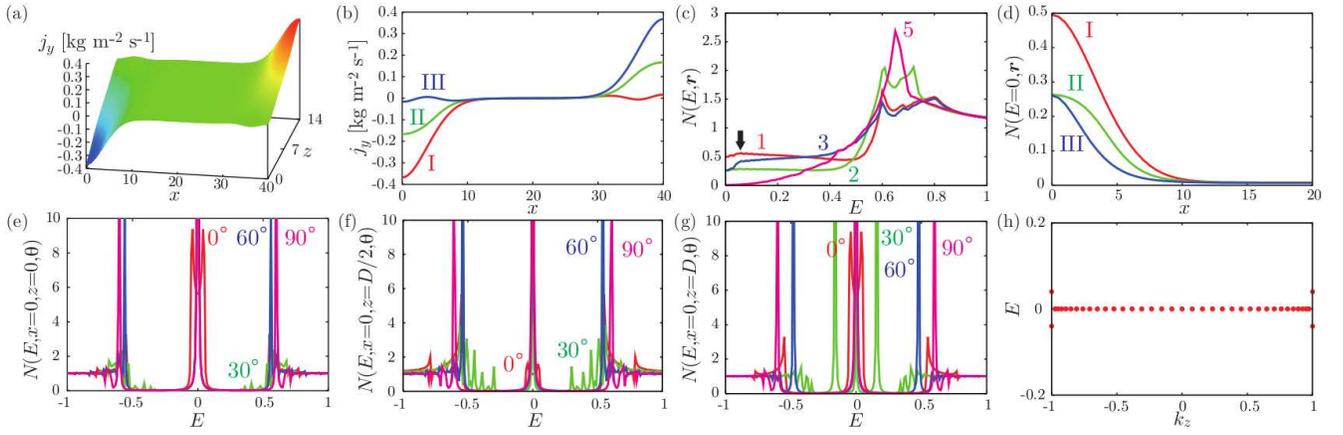}
\end{center}
\caption{(Color online) 
Calculated results for the A-phase with D=14.
(a) Mass current $j_y(\bi{r})$ in the $x$-$z$ plane.
(b) Profiles of the mass current $j_y(\bi{r})$ along paths ``I", ``II", and ``III".
(c) LDOS $N(E,\bi{r})$ at ``1", ``2", ``3", and ``5".
(d) Zero-energy LDOS $N(E=0,\bi{r})$ along ``I", ``II", and ``III".
Angle-resolved LDOS $N(E,\bi{r},\theta)$ with $\phi=0^{\circ }$ at ``1" (e), ``2" (f), and ``3" (g).
(h) The peak energy of angle-resolved LDOS as a function of $k_z$ for $k_y=0$ at ``1".}
\label{Al}
\end{figure*}

The mass current $j_y(\bi{r})$ is shown in Fig.~\ref{Al}(a),
and its profiles along ``I" ($z=0$), ``II" ($z=D/2$), and ``III" ($z=D$) are shown in Fig.~\ref{Al}(b).
The mass current flows clockwise relative the local $l$-vector direction.
The $j_y$ from $l$-vector in the bulk region flows to the negative (positive) $y$-direction at $x=0$ ($x=L$).
At the side edges, since $l$-vector points to the negative $x$-direction,
the $j_y$ from the edges flows to the negative and positive $y$-directions in $z<D/2$ and $z>D/2$, respectively.
Therefore, the edge mass current is strengthened in $z<D/2$ ($z>D/2$) and weakened in $z>D/2$ ($z<D/2$) at the edge $x=0$ ($x=L$).
The symmetry of the mass current between $z<D/2$ and $z>D/2$ is broken despite the symmetry of the OP.

Figure~\ref{Al}(c) shows LDOS at $(x,z)$ $=$ $(0,0)$, $(0,D/2)$, $(0,D)$, and $(L/2,D/2)$.
At $(0,0)$ and $(0,D)$ (the lines ``1" and ``3"), 
there are the zero-energy LDOS and a small peak at a low energy indicated by the arrow in Fig.~\ref{Al}(c). 
Although $l$-vector points to the edge, the zero-energy LDOS also exists at $(0,D/2)$ (the line ``2").
The lines ``1" and ``3" are slightly decreased and increased, respectively, in low energy from the small peak as energy increases.
The line ``2" is almost constant in low energy.
The LDOS (the lines ``1", ``2", and ``3") are enhanced sharply near $E\approx 0.5$.
The LDOS at $(L/2,D/2)$ (the line ``5") shows the typical spectral behavior of the point node.
In Fig.~\ref{Al}(d) we show the extent of the zero-energy edge mode toward the bulk from the edge at $x=0$.
The amounts of the zero-energy LDOS at $x=0$ along $z=D/2$ and $z=D$ (the lines ``II" and ``III") 
are smaller than that along $z=0$ (the line ``I").
The extent of the zero-energy LDOS along ``III" is of the order of $5\xi_0$ as in the thin slab $D=8$; 
however, the zero-energy LDOS along ``I" and ``II" are more widely extended.

The angle-resolved LDOS $N(E,\bi{r},\theta)$ with $\phi=0^{\circ }$ at $(0,0)$, $(0,D/2)$, and $(0,D)$ 
are, respectively, displayed in Figs.~\ref{Al}(e), \ref{Al}(f), and \ref{Al}(g).
The zero-energy peaks of the angle-resolved LDOS appear except for $\theta=0^{\circ }$.
The peaks of the line $\theta=0^{\circ }$ at the finite low energy appear from the upper and lower surfaces
because the $k_z$-component of the pair potential is non-vanishing.
Since thickness $D=14$ is small, 
the original zero-energy peak of $\theta=0^{\circ }$ perpendicular to the upper and lower surfaces is split into the finite energy peaks,
which is discussed particularly for the B-phase in next section.
The splitting peaks at $(0,0)$ are shown in Fig.~\ref{Al}(h) only at $k_z=\pm 1$ for $k_y=0$.
The finite energy peaks compose the small peak of the LDOS in the lines ``1" and ``3" in Fig.~\ref{Al}(c).
Since the position $(0,D/2)$ is apart form the upper and lower surfaces, the energy peaks of $\theta=0^{\circ }$ are small.
Consequently, the peak of the LDOS at low energy is not clear (the line ``2" in Fig.~\ref{Al}(c)).
The point nodes of the OP near the side edges have a tilt to the $x$-direction according to $l$-vector,
or the anti-nodes of the OP lie to some angles from the $k_z$-axis.
Thus, the incident QPs from $(0,0)$ with  low angles from the $k_z$-axis
do not feel the clear small gap in the vicinity of the point nodes.
For instance, the line $\theta=30^{\circ }$ in Fig.~\ref{Al}(e) should be compared with that in Fig. \ref{As}(e) for the thin slab $D=8$.
Therefore, the gap-like LDOS enhanced sharply near $E\approx 0.5$ appears (the line ``1" in Fig.~\ref{Al}(c)).
On the other hand, the incident QPs from $(0,D)$ with low angles from the $k_z$-axis are reflected at the edge
and regarded as with low angles from the $-k_z$-axis.
Since the QPs feel the small gap in the vicinity of the point nodes,
they have the clear peaks in the angle-resolved LDOS (Fig.~\ref{Al}(g)).
Therefore, the LDOS is slightly increased in low energy as energy increases (the line ``3" in Fig.~\ref{Al}(c)).
Since the QPs at $(0,D/2)$ have the two kinds of the characteristic momentum,
the LDOS is almost constant in low energy (the line ``2" in Fig.~\ref{Al}(c)).

The LDOS at the side edge is different and depends on the thickness of the slab.
In the thin slab ($D=8$), $l$-vector points to the $z$-direction everywhere and the LDOS is $N(E)=N(0)+\alpha E^2$ in low energy.
In contrast, in the thick slab ($D=14$), 
$l$-vector points to the $x$-direction near the side edge and the LDOS is $N(E)=N(0)+N_{\rm gap}(E)$,
where $N_{\rm gap}(E)$ is almost zero in low energy and increases sharply near $E\approx 0.5$ like the LDOS with the full gap state.
The edge mass current varies along the side edge in the thick slab, which is also a great difference with the thin slab.

Note that the A-phase is metastable in the slab with $D=14$ at $T=0.2T_c$ 
in the weak-coupling limit for $P\rightarrow 0$.~\cite{vorontsov:2007}
In the same slab, the A-phase is stabilized at $T=0.9T_c$;~\cite{vorontsov:2007}
however, $l$-vector points to only the $z$-direction,
because the $k_z$-component is absent owing to the longer coherence length at higher temperature.
The mechanism is similar to that the A-phase is more stable than the B-phase, which has the $k_x$-, $k_y$-, and $k_z$-components, 
in a thin slab at low temperature.
The calculated results shown in this subsection is not in free energy minimum at $P=0$; however,
the strong-coupling effect by pressure stabilizes the A-phase texture.
Thus, the above characteristics from the texture will be observed at finite pressure.

\section{\label{sec:B}B-phase}

The B-phase is stable when the thickness of a slab is larger than $\approx 13\xi_0$ at $T=0.2T_c$.~\cite{vorontsov:2007}
The OP of the B-phase is described by~\cite{vollhardt:book}
\begin{align*}
A_{\mu i}=R_{\mu i}(\bi{n}_d,\theta_d)
\begin{pmatrix}
A_{xx} & 0 & 0 \\
0 & A_{yy} & 0 \\
0 & 0 & A_{zz}
 \end{pmatrix},
\end{align*}
where $R_{\mu i}(\bi{n}_d,\theta_d)$ is a rotation matrix with a rotation axis $\bi{n}_d$ and a rotation angle $\theta_d$ about $\bi{n}_d$.
The rotation matrix gives the relative angle between the orbital momentum and the direction along which the spin of a Cooper pair is zero.
The spin state is stable by the dipole-dipole interaction when $\theta_d=\theta_L \equiv \cos^{-1}(-1/4)$ 
and $\bi{n}_d$ is perpendicular to the surface in the absence of a magnetic field.~\cite{vollhardt:book}
If the thickness of a slab is much smaller than the dipole coherence length, $\bi{n}_d$ is locked to the $z$-axis.
Thus, we derive $A_{xx}(\bi{r})$, $A_{yy}(\bi{r})$, and $A_{zz}(\bi{r})$
with the uniform rotation matrix $R(\bi{z},\theta_L)$.
Note that the three components have the same phase.

At the surface of a slab, since the normal component to the surface of the orbital state vanishes, the planar state will be realized.
The OP in the planar state is described by~\cite{vollhardt:book}
\begin{align*}
A_{\mu i}=R_{\mu i}(\bi{n}_d,\theta_d)
\begin{pmatrix}
A_{xx} & 0 & 0 \\
0 & A_{yy} & 0 \\
0 & 0 &0
 \end{pmatrix},
\end{align*}
where we regard the $z$-axis as perpendicular to the edge.
If $\theta_d=0$, the OP can be written by other descriptions as
\begin{align*}
\Delta_{\uparrow \uparrow }(\bi{k})=-\frac{\Delta_p}{\sqrt{2}}(k_x-ik_y),\ \Delta_{\downarrow \downarrow }(\bi{k})=\frac{\Delta_p}{\sqrt{2}}(k_x+ik_y),
\end{align*}
where $\Delta_{\uparrow \uparrow }=-(\Delta_x-i\Delta_y)/\sqrt{2}$, $\Delta_{\downarrow \downarrow }=(\Delta_x+i\Delta_y)/\sqrt{2}$,
and $\Delta_p=A_{xx}=A_{yy}$.
Therefore, in the planar state, $S_z=+1$ spin state has $L_z=-1$ orbital angular momentum and $S_z=-1$ spin state has $L_z=+1$ orbital angular momentum.

\subsection{$D=30\xi_0$: Thick slab for B-phase}

\begin{figure}
\begin{center}
\includegraphics[width=8.5cm]{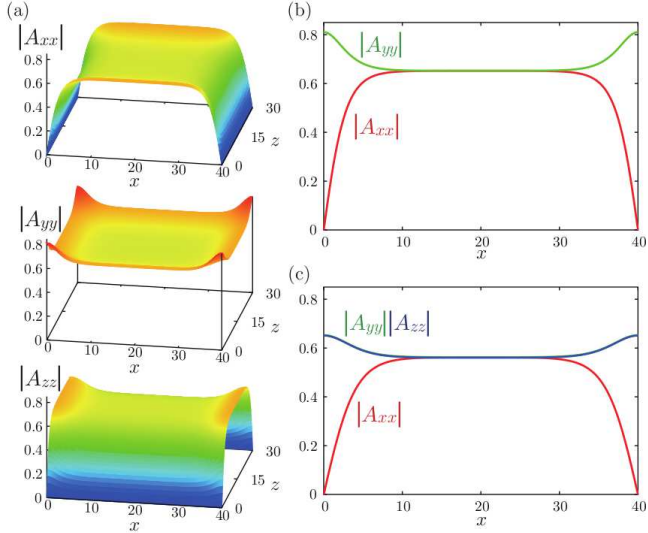}
\end{center}
\caption{(Color online) 
The OP for the B-phase with D=30.
(a) Amplitude of each component.
Profiles of each component along ``I" (b), and along``II" (c).}
\label{B-OP}
\end{figure}

We consider that the thickness of a slab is much longer than the coherence length, namely, the thickness can be regarded as macroscopic for the OP, 
and much shorter than the dipole coherence length.
The requirement is satisfied for a thickness of $D=30$.
It is clearly seen that the component of the OP perpendicular to the edge becomes zero in Fig.~\ref{B-OP}(a).
The polar state $k_y$ occurs at the corner ``1" ($x=0,z=0$).
The planar state $k_x-ik_y$ for the up-up spin Cooper pairs 
and $k_x+ik_y$ for the down-down spin Cooper pairs is attained in the middle region around ``4" ($x=L/2,z=0$), shown in Fig.~\ref{B-OP}(b).
The planar state $k_y\pm ik_z$ is realized at the side edge ``2" ($x=0,z=D/2$)
and the OP is recovered to the bulk value of the B-phase around ``5" ($x=L/2,z=D/2$), shown in Fig.~\ref{B-OP}(c).
At all edges ($x=0$, $x=L$, $z=0$, and $z=D$), the Majorana fermion edge state exists.~\cite{chung:2009, volovik:2009a, nagato:2009}

\begin{figure}
\begin{center}
\includegraphics[width=8.5cm]{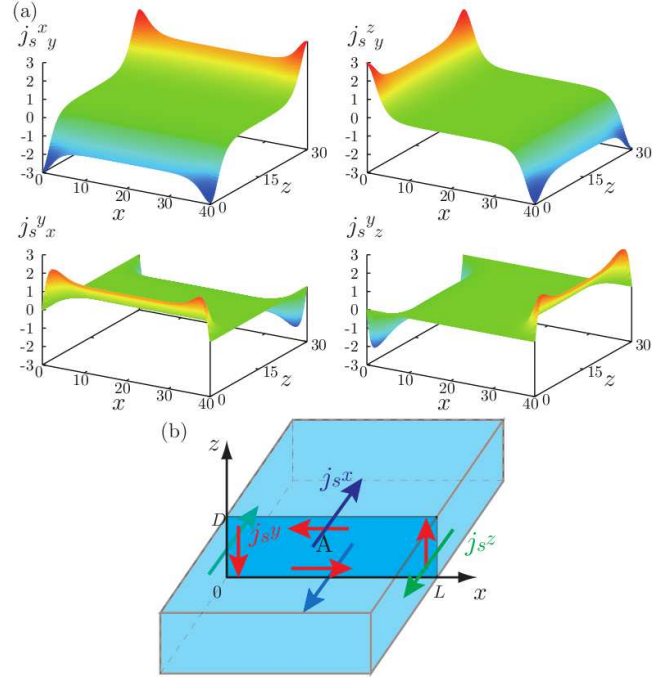}
\end{center}
\caption{(Color online) 
(a) The spin current for the B-phase with $D=30$.
The unit of the spin current is $10^{-9}$ ${\rm J}/{\rm m}^2$.
(b) A schematic flow of the spin current for $S_x$, $S_y$, and $S_z$ components.} 
\label{B-c}
\end{figure}

The spin current $j_{sy}^x(\bi{r})$, $j_{sy}^z(\bi{r})$, $j_{sx}^y(\bi{r})$, and $j_{sz}^y(\bi{r})$ are shown in Fig.~\ref{B-c}(a).
Since the spin current flows in the three-dimensional real space, we show the schematic flow of the spin current  in Fig.~\ref{B-c}(b).
The spin current for the $i$-component of spin turns around the $i$-axis.
This is understandable as follows:
Since the up-up spin Cooper pairs have negative chirality and the down-down spin Cooper pairs have positive chirality,
the $z$-component of the spin current turns around the $z$-axis.
Other components are also understood in the same manner.
Note that the schematic flow in Fig.~\ref{B-c}(b) is derived from $A_{ii}$ before rotation by $R(\bi{z},\theta_L)$.
The rotation axes for the $x$- and $y$-component of the spin current are changed correspondingly.

\begin{figure}
\begin{center}
\includegraphics[width=8.5cm]{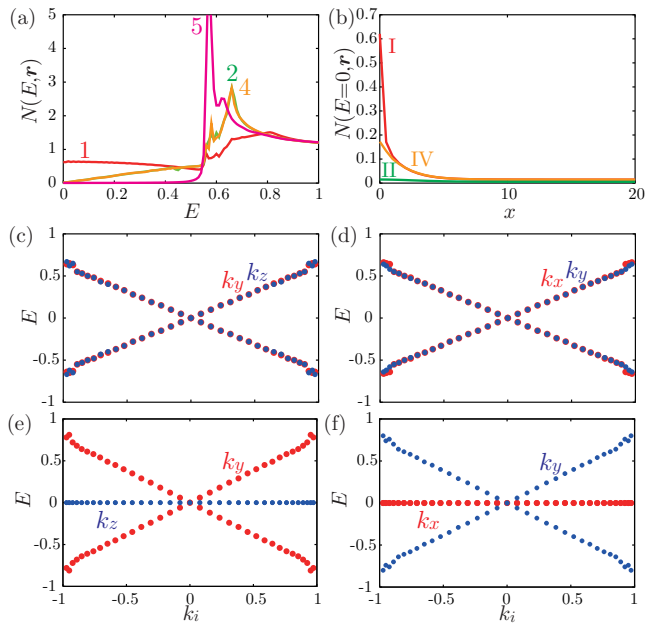}
\end{center}
\caption{(Color online) 
The LDOS for the B-phase with $D=30$.
(a) LDOS $N(E,\bi{r})$ at ``1", ``2", ``4", and ``5".
(b) Zero-energy LDOS $N(E=0,\bi{r})$ along ``I", ``II", and ``IV" ($z=0.5$).
The peak energy of angle-resolved LDOS as a function of $k_y$ for $k_z=0$ and as a function of $k_z$ for $k_y=0$ at ``2" (c), 
and as a function of $k_x$ for $k_y=0$ and as a function of $k_y$ for $k_x=0$ at ``4" (d).
(e) and (f) are the same plots as (c) and (d), but at ``1".}
\label{Bl}
\end{figure}

Figure~\ref{Bl}(a) shows the LDOS.
It is seen that only the line ``1" ($x=0,z=0$) has the zero-energy LDOS 
from the zero-energy Andreev bound state of QPs without the $k_y$-component.
The lines ``2" and ``4" ($x=0,z=D/2$ and $x=L/2,z=0$) are almost the same
and show the linear relation $N(E,\bi{r})\propto E$ near $E=0$ which reflects the surface Andreev bound states.~\cite{buchholtz:1981,nagato:1998}
The line ``5" ($x=L/2,z=D/2$) corresponds to the bulk LDOS where the full gap is expected.
In Fig.~\ref{Bl}(b) we show the extension of the zero-energy state toward the middle from the edge at $x=0$.
The zero-energy LDOS of the line ``I" ($z=0$) reduces sharply as approaching the middle region from the edge.
It is found that the value $N(E=0,x=0)$ in the line ``IV" ($z=0.5$) is much smaller than that of the line ``I".
The line ``II" along $z=D/2$ shows the absence of the zero-energy LDOS.
Therefore, we conclude that the zero-energy LDOS is localized at the corner of the order of $\xi_0$.
The peak energy of the angle-resolved LDOS is shown in Figs.~\ref{Bl}(c) for the side edge and \ref{Bl}(d) for the lower surface.
Figure~\ref{Bl}(c) is as a function of $k_y$ for $k_z=0$ and as a function of $k_z$ for $k_y=0$ at ``2" ($x=0,z=D/2$).
Figure~\ref{Bl}(d) is as a function of $k_x$ for $k_y=0$ and as a function of $k_y$ for $k_x=0$ at ``4" ($x=L/2,z=0$).
These show that the side edge and lower surface have the same dispersion relation for low energy.
The linear dispersion forms the Majorana cone at the surface.~\cite{chung:2009}

The peak energy of the angle-resolved LDOS at the corner $(0,0)$
is shown in Figs.~\ref{Bl}(e) and \ref{Bl}(f).
The energy is dispersionless along the $k_z$- and $k_x$-axes for $k_y=0$.
The Dirac valley-type dispersion is composed of the zero-energy LDOS at the corner (the line ``1" in Fig.~\ref{Bl}(a)).
This is different from the Majorana cone at the surface.

\subsection{$D=14\xi_0$: Thin slab for B-phase}

We consider $D=14$ in thickness of a slab which is the critical thickness for stabilizing the B-phase
below which the stripe phase becomes stable.~\cite{vorontsov:2007}
Since the thickness is short, the $k_z$-component of the OP can not completely recover the bulk value.
Except that the $k_z$-component is reduced, the qualitative features of the OP and the spin current are the same in the case of $D=30$;
however, the LDOS shows different spectrum.

\begin{figure}
\begin{center}
\includegraphics[width=8.5cm]{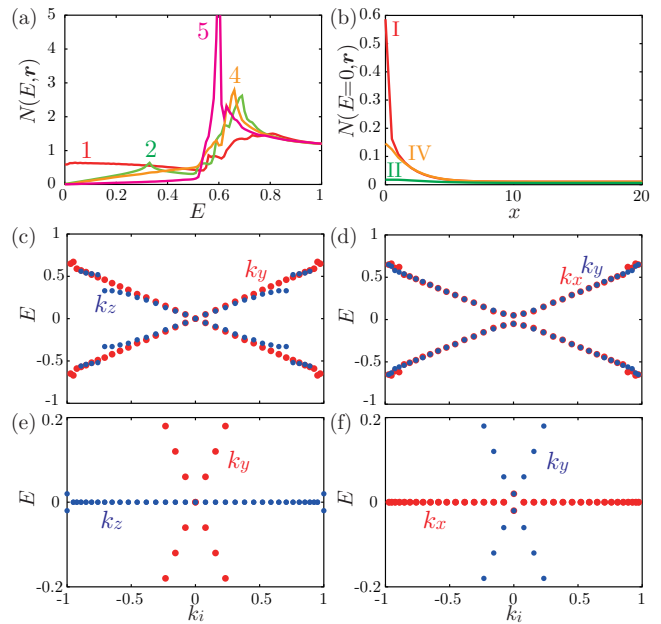}
\end{center}
\caption{(Color online)
The same as in Fig.~\ref{Bl}, but for the B-phase with $D=14$.}
\label{Bm}
\end{figure}

Figure~\ref{Bm} shows the LDOS and the peak energy of the angle-resolved LDOS, which are compared with those in Fig.~\ref{Bl}.
The differences between them are following:
Since the thickness of the slab is short, 
the line ``5" for the center ($x=L/2,z=D/2$) has the spectrum from the full gap
and the linear relation near $E=0$, shown  in Fig.~\ref{Bm}(a).
The latter comes from the contributions extending from the upper and lower surfaces.
The line ``2" for the side edge ($x=0,z=D/2$) in Fig.~\ref{Bm}(a) has the small peak at $E\approx 0.35$
composed of the deformed Majorana cone in $0.5\lessapprox |k_z| \lessapprox 0.7$, shown in Fig.~\ref{Bm}(c),
because the QPs with the $k_z$-component reflect that the $k_z$-component of the OP does not recover the bulk value.
The peak energy at ``4" for the lower surface in Fig.~\ref{Bm}(d) implies that the lowest energy of the Majorana cone is lifted from zero 
because of the tunneling of the zero-energy modes bound at two surfaces corresponding to the upper and lower surfaces.
For the same reason, 
the zero-energy modes at the corner is split for $k_z=\pm 1$ in Fig.~\ref{Bm}(e) and for $k_y=0$ or $k_x=0$ in Fig.~\ref{Bm}(f).
By the energy splitting, the zero-energy LDOS is slightly reduced;
however, the extension of the zero-energy state is the same for the thick slab $D=30$, shown in Fig.~\ref{Bm}(b).

\subsection{Dependence of the energy splitting on the thickness}

\begin{figure}
\begin{center}
\includegraphics[width=8.5cm]{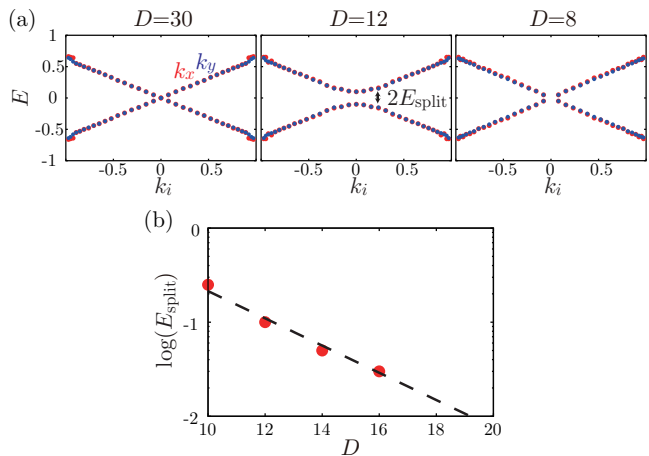}
\end{center}
\caption{(Color online)
Calculated results for the B-phase at ``4".
(a) The peak energy of angle-resolved LDOS as a function of $k_x$ for $k_y=0$ and as a function of $k_y$ for $k_x=0$ with several thicknesses
$D=30$, $12$, and $8$.
(b) The dependence of the energy splitting on the thickness.
The dashed line is proportional to $\exp(-D/3\xi_0)$. }
\label{split}
\end{figure}

The lowest energy of the Majorana cone at the lower surface is more lifted from the zero-energy as the thickness is shorter.
If the thickness is macroscopically long, the zero-energy modes appear, 
shown in the left panel of  Fig.~\ref{split}(a) for the B-phase at ``4" ($x=L/2,z=0$) with $D=30$.
When the thickness is short, the zero-energy modes bound at two surfaces are hybridized with each other.
They form the symmetric and anti-symmetric states with the opposite sign energy $E_+$ and $E_-$, respectively, where $|E_+|=|E_-|=E_{\rm split}$.
The representative dispersion relation with $D=12$ is shown in the middle panel in Fig.~\ref{split}(a).
In the case of more shorter thickness, the B-phase becomes the planar phase, which is metastable against the A-phase,
because the $k_z$-component of the OP vanishes.
Since the Fermi surface has the point nodes toward the direction of $k_x=k_y=0$,
the angle-resolved LDOS toward the direction corresponds to that of the normal state without the peak energy.
We show the dispersion relation with $D=8$ except the point $k_x=k_y=0$ in the right panel in Fig~\ref{split}(a).
There is a difference between the dispersion relations for $D=12$ and $D=8$ also in long wavelength
which are curved and linear, respectively.

The dependence of the energy splitting on the thickness is shown in Fig.~\ref{split}(b).
The energy splitting has the relation $E_{\rm split}\propto \exp(-D/3\xi_0)$,
where $3\xi_0$ is effective coherence length.
The exponential suppression is similar to 
the energy splitting of the Majorana zero-energy modes bound at two vortices in a chiral $p$-wave superfluid.~\cite{cheng:2009,mizushima:2010b}

\section{\label{sec:discuss}Discussion}

\subsection{Majorana zero modes in the A- and B-phases}

In the A-phase which has point nodes,
since the zero-energy QPs appear on the line $k_y=0$ in the momentum space,
the zero-energy QPs are dispersionless along the $k_z$-direction (Fig.~\ref{As}(g)).
Therefore, the dispersion of the QPs forms the Dirac valley
and the zero-energy LDOS spreads over the order of $5\xi_0$ from the side edge (Figs.~\ref{As}(c) and \ref{As}(d)).
In contrast, since the OP in the B-phase has full gap on the Fermi surface, only the QPs with the momentum perpendicular to the edge
are the zero-energy Majorana QPs.
Therefore, the dispersion of the QPs forms the Majorana cone (Figs.~\ref{Bl}(c) and \ref{Bl}(d)).
However, at the corner, the dispersion of the QPs forms the Dirac valley and the zero-energy LDOS spreads over the order of $\xi_0$.
Types of dispersion relation for the A- and B-phases are summarized in Table~\ref{tab}.

\begin{table*}
\begin{center}
\begin{minipage}{15cm}
\caption{Types of dispersion relation at the upper or lower surface, side edge, and corner of a slab for the A- and B-phases.
Dirac valley is formed from linear dispersion from $E=0$ along a certain momentum direction 
and dispersionless zero-energy modes along another momentum direction in long wavelength.
Majorana cone is formed from isotropic liner dispersion from $E=0$ on the plane parallel to a surface in the momentum space in long wavelength,
where the zero mode is in a point on the Fermi surface.
The mark $\times$ indicates absence of zero-energy modes.}
\label{tab}
\end{minipage}
\begin{tabular*}{15cm}{@{\extracolsep{\fill}}ccccc}
\hline \hline
\rule{0pt}{2.5ex} & & upper or lower surface & side edge & corner \\ \hline
\rule{0pt}{2.5ex}A-phase & thin slab ($D=8\xi_0$) & $\times$ & Dirac valley & Dirac valley \\
\rule{0pt}{2.5ex} & thick slab ($D=14\xi_0$) &$\times$ &  Dirac valley & Dirac valley \\ \hline
\rule{0pt}{2.5ex}B-phase & thin slab ($D=14\xi_0$) & $\times$ & Majorana cone & Dirac valley \\
\rule{0pt}{2.5ex} & thick slab ($D=30\xi_0$) & Majorana cone & Majorana cone & Dirac valley \\ \hline \hline
\end{tabular*}
\end{center}
\end{table*}

The spin degeneracy of the QPs is also different between the A- and B-phases.
In the A-phase, since the up-up and down-down spin Cooper pairs have the same chirality,
the low energy QPs have the degenerate branch of the dispersion at the edge (Fig.~\ref{As}(h)).
In the B-phase, since the up-up and down-down spin Cooper pairs have the opposite chirality,
the low energy QPs have two branches of the dispersion at the edge (Fig.~\ref{Bl}(c)).
This difference manifests itself in the edge current, namely,
the mass current in the A-phase (Fig.~\ref{As}(b)) and the spin current in the B-phase (Fig.~\ref{B-c}).

\subsection{Stripe phase}

We discuss the Majorana zero modes at a domain wall in the stripe phase.~\cite{vorontsov:2007}
We take the thickness of a film along the $z$-direction and the modulation of the OP along the $x$-direction.
In the stripe phase, the $k_z$-component of the OP changes the sign at the domain wall perpendicular to the $x$-direction
so that the pair breaking by the reflection at the surface of the film is prevented.
Then, the OP is described as $\bi{\Delta }_{\rm right}=(\Delta_{\parallel }k_x,\Delta_{\parallel }k_y,\Delta_{\perp }k_z)$ to the right of the domain wall
and $\bi{\Delta }_{\rm left}=(\Delta_{\parallel }k_x,\Delta_{\parallel }k_y,-\Delta_{\perp }k_z)$ to the left of the domain wall,
where $\Delta_{\parallel }$ is finite everywhere and $\Delta_{\perp }$ vanishes at the domain wall.

The QPs with the $k_z$-component across the domain wall have the finite energy Andreev bound states
because they feel the sign change of the $k_z$-component of the pair potential, which is not the exact $\pi$-phase shift.
In addition, since the QPs with $k_z=0$ feel the full gap of the pair potential, they are not excited in low energy.
Therefore, the Majorana zero-energy QP is absent at the domain wall in the stripe phase.
The domain wall is qualitatively different from the edge.

\subsection{Experimental proposal}

There are several experimental means to detect the Majorana nature.
Surface specific heat measurement, 
which was performed in connection with detection of the Andreev surface bound state,~\cite{choi:2006}
resolves the side edge contribution $C_{\rm surface}(T) = \gamma T$ of the A-phase in a thin slab at low temperatures,
where $\gamma \propto N(E= 0)$,
because the bulk contribution $C_{\rm bulk} \propto T^3$ 
which comes from point nodes where $N(E) \propto E^2$ is distinguishable.
Note that, in the A-phase, the contribution from the two upper and lower specular surfaces in the slab geometry is the same as that from the bulk.
Thus, the surface specific heat $C_{\rm surface}(T) = \gamma T$ of the Majorana QPs is distinctive.
If the $l$-vector direction is modulated near the side edge,
the LDOS at the surface is $N(E)=N(0)+N_{\rm gap}(E)$.
The surface specific heat from the LDOS is also $C_{\rm surface}(T) = \gamma T$.

In the B-phase, the zero-energy LDOS is localized at the corner of the order of $\xi_0$.
The contribution from the corner is interesting but smaller than that of the surface.
Since we will discriminate $C_{\rm bulk}\propto T^{-3/2}e^{-\Delta/k_BT}$ from the gap $\Delta$ 
and $C_{\rm surface}\propto T^2$ from the linear behavior of LDOS $N(E,\bi{r})\propto E$ near $E=0$,
the Majorana fermion can be observed.
In the slab with short thickness where the zero-energy modes are split at the upper and lower surfaces,
the difference of the specific heat from the surface and bulk will not be distinctive
because the gap structure and linear behavior of the LDOS coexist.

The observation of the edge mass current in the A-phase, which is intimately connected with the intrinsic angular momentum,~\cite{vollhardt:book}
is also hopeful.
The magnitude of the edge mass current is unchanged in wider slabs than $L=40\xi_0$ for which we have calculated.
Considering the observation of the torque from the edge mass current for a 10 mm $\times$ 7 mm $\times$ 0.6 $\mu$m slab sample (Bennett {\it et al.}~\cite{bennett:2010} had been used) by a typical torsional oscillator,
the frequency shift from the edge mass current is of the order of $10^{-23}$ Hz.~\cite{okuda:private}
The torque is too small to observe by a torsional oscillator because the magnitude is $\sim N\hbar$,
where $N$ is the total number of $^3$He atoms in the slab sample.
We have to consider other experimental methods.
Also the edge spin current in the B-phase which flows three-dimensionally has been obtained quantitatively.
The techniques to detect the spin current is desired.
The specific experimental proposal to observe the edge current is a future problem.

The most direct evidence of the Majorana nature is derived from the observation of the anisotropic spin susceptibility. 
If we use the Majorana nature of the edge state, in the A-phase,
the local spin operators result in $S_x \approx S_y \approx 0$ 
and only $S_z$ parallel to $d$-vector remains nontrivial for $T\ll T_c$.~\cite{stone:2004}
This predicts the Ising-like spin dynamics for the local spin operator parallel to $d$-vector in the A-phase as well as
that perpendicular to the edge in the B-phase.~\cite{chung:2009,nagato:2009}
This is in sharp contrast to the susceptibility parallel to $d$-vector in the bulk A-phase
which is suppressed at low temperatures according to the Yosida function.
On the other hand, the susceptibility perpendicular to $d$-vector still assumes the bulk value, which is the same as it in the normal state.
The anisotropic susceptibility has been discussed also by Shindou {\it et al.}~\cite{shindou:2010}
and has been calculated in the B-phase by Nagato {\it et al.}~\cite{nagato:2009}

QP scattering  or QP beam experiments are extremely interesting.
They were performed in the past on $^4$He where roton-roton scattering is
treated~\cite{forbes:1990} and on the $^3$He B-phase where the surface Andreev
bound state is investigated.~\cite{enrico:1993,okuda:1998}
Using this method, we may pick up
Majorana QPs with a particular wave number.
Particularly in the A-phase, the Majorana QPs from the edge is separated from other QPs from the nodal region.

Another option might be to use a free surface where the Majorana fermion surface state is formed.
As shown by Kono,~\cite{kono:2010} it can be detected through the 
excitation modes of the floating Wigner lattice of electrons placed on the surface.
We need a special, but feasible configuration of the experimental setups.

Note that the recent work of transverse acoustic impedance measurements to detect the surface bound states in the superfluid $^3$He
will derive the important information of the Majorana QPs.~\cite{murakawa:2009,murakawa:2010}

\section{Summary}

We have designed a concrete experimental setup to observe the Majorana nature at the surface in the slab geometry. 
In connection with realistic slab samples,
we have considered the upper and lower surfaces and the side edges including the corners with several thicknesses.
We have demonstrated that the quasi-classical Eilenberger equation yields the quantitatively reliable information 
on physical quantities for the superfluid $^3$He A- and B-phases.
Specifically, we have exhibited the difference of LDOS between the A- and B-phases
and evaluated the mass current for the A-phase and the spin current for the B-phase quantitatively.
Then, we have shown the influence on the Majorana zero modes from the spatial variation of $l$-vector for the A-phase in the thick slab
and the energy splitting of the zero-energy modes for the B-phase confined in the thin slabs.
The corner of the slab in the B-phase is accompanied by the unique zero-energy LDOS of corner modes.
In addition, we have demonstrated the absence of the Majorana zero-energy QP at the domain wall in the stripe phase.
On the basis of the quantitative consequences, it is proposed that the measurement of the specific heat, the edge current, 
and the anisotropic spin susceptibility provides feasible and verifiable experiments to check the Majorana nature.
The control on the thickness of the slab is crucial to detect the Majorana surface states.
The experiment controlling the thickness of the film of the superfluid $^3$He is interesting.~\cite{saitoh:2007}

\begin{acknowledgements}

We thank T. Mizushima for helpful theoretical discussions and
K. Kono, J. Saunders, and Y. Okuda for informative discussions on their experiments.
Y.~T. acknowledges the support of the Research Fellowships of the Japan Society for the Promotion of Science for Young Scientists.

\end{acknowledgements}

\appendix*
\section{}

We use the symmetry of the quasi-classical Green's function in Eilenberger Eq.~\eqref{Eilenberger eq} to reduce computational time.
If we replace Matsubara frequency $\omega_n$ with $-\omega_n^*$,
the quasi-classical Green's functions in particle-hole space have relations
\begin{align}
&\hat{g}(\bi{k},\bi{r},-\omega_n^*) = -\hat{g}(\bi{k},\bi{r},\omega_n)^{\dagger },\nn\\
&\underline{\hat{g}}(\bi{k},\bi{r},-\omega_n^*) = -\underline{\hat{g}}(\bi{k},\bi{r},\omega_n)^{\dagger }, \nn\\
&\hat{f}(\bi{k},\bi{r},-\omega_n^*) = \underline{\hat{f}}(\bi{k},\bi{r},\omega_n)^{\dagger },\nn\\
&\underline{\hat{f}}(\bi{k},\bi{r},-\omega_n^*) = \hat{f}(\bi{k},\bi{r},\omega_n)^{\dagger },
\label{omega symmetry}
\end{align}
where we describe the complex conjugate of the Matsubara frequency explicitly 
because that is important when we calculate LDOS.
By the relations, we are allowed to sum only the positive $\omega_n$ to calculate self-consistent pair potential
and mass and spin currents.

If we reverse the sign of relative momentum $\bi{k}$,
the sign of spin-triplet components of the OP changes; on the other hand, that of a spin singlet component of the OP does not change.
Specifically, general OP
\begin{align*}
\hat{\Delta }(\bi{k},\bi{r}) =
\begin{pmatrix}
\Delta_{\uparrow\uparrow }(\bi{k},\bi{r}) & \Delta_{\uparrow\downarrow }(\bi{k},\bi{r}) \\
\Delta_{\downarrow\uparrow }(\bi{k},\bi{r}) & \Delta_{\downarrow\downarrow }(\bi{k},\bi{r})
\end{pmatrix}
\end{align*}
has a relation, $\hat{\Delta}(-\bi{k},\bi{r})=-\hat{\Delta}^{\rm T}(\bi{k},\bi{r})$,
where a superscript T indicates transposition of a matrix.
Since the sign of Fermi velocity also changes, $\bi{v}(-\bi{k})=-\bi{v}(\bi{k})$,
the quasi-classical Green's functions in particle-hole space have relations
\begin{align}
&\hat{g}(-\bi{k},\bi{r},\omega_n^*) = \underline{\hat{g}}(\bi{k},\bi{r},\omega_n)^*,\nn\\
&\underline{\hat{g}}(-\bi{k},\bi{r},\omega_n^*) = \hat{g}(\bi{k},\bi{r},\omega_n)^*, \nn\\
&\hat{f}(-\bi{k},\bi{r},\omega_n^*) = -\underline{\hat{f}}(\bi{k},\bi{r},\omega_n)^*,\nn\\
&\underline{\hat{f}}(-\bi{k},\bi{r},\omega_n^*) = -\hat{f}(\bi{k},\bi{r},\omega_n)^*.
\label{k symmetry1}
\end{align}

More reduction of computational time is possible by using mirror operators which define 
$S_x\bi{a}\equiv(-a_x,a_y,a_z)$, $S_z\bi{a}\equiv(a_x,a_y,-a_z)$, and $S_{xz}\bi{a}\equiv(-a_x,a_y,-a_z)$
with an arbitrary vector $\bi{a}=(a_x,a_y,a_z)$.
If the mirror operators act on the center-of-mass coordinate of pair potential 
under the antiperiodic boundary condition mentioned in Sec.~\ref{sec:sys}, 
$\Delta(\bi{k},S\bi{r})=\Delta(S\bi{k},\bi{r})$, where $S$ is one among $S_x$, $S_z$, and $S_{xz}$.
Since $S\bi{v}\cdot S\bi{\nabla}=\bi{v}\cdot \bi{\nabla}$, the quasi-classical Green's function satisfies
\begin{align}
\widehat{g}(S\bi{k},S\bi{r},\omega_n) = \widehat{g}(\bi{k},\bi{r},\omega_n).
\label{k symmetry2}
\end{align}

Therefore, the quasi-classical Green's function and pair potential are obtained self-consistently by numerical calculation
for only the positive Matsubara frequency and  one eighth of the Fermi surface in the coordinate $-L/2 \le x \le L/2$ and $-D/2 \le z \le D/2$
by the symmetry of the quasi-classical Green's function 
\eqref{omega symmetry}, \eqref{k symmetry1}, and \eqref{k symmetry2}.
The calculation for the real space coordinates is carried out by parallel computing with OpenMP.

%\bibliographystyle{apsrev4-1}
%\bibliography{references}

%

\end{document}